# Carrier Doping Modulates 2D Intrinsic Ferromagnetic $Mn_2Ge_2Te_6$ Monolayer High Curie Temperature, Large Magnetic Crystal Anisotropy

Ziyuan An,[†] Ya Su,[ξ] Shuang Ni,[#*] Zhaoyong Guan[†§*]

[†]Key Laboratory of Colloid and Interface Chemistry, Ministry of Education, School of Chemistry and Chemical Engineering, Shandong University, Jinan, Shandong 250100, P. R. China

[ξ] School of Electrical Engineering, Shandong University, Jinan, Shandong 250100, P. R. China

[#]Research Center of Laser Fusion, China Academy of Engineering Physics. Mianyang, Sichuan 621900, P. R. China

[§]Science Center for Material Creation and Energy Conversion Institute of Frontier and Interdisciplinary Science School of Chemistry and Chemical Engineering, Shandong University, Jinan, Shandong 250100, P. R. China

## ABSTRACT

The $Mn_2Ge_2Te_6$ shows intrinsic ferromagnetic (FM) order, with Curie temperature ($T_c$) of 316 K. The FM order origins from superexchange interaction between Mn and Te atoms. $Mn_2Ge_2Te_6$ is half-metal (HM), and spin-β electron is a semiconductor with gap of 1.462 eV. $Mn_2Ge_2Te_6$ tends in-plane anisotropy (IPA), with magnetic anisotropy energy (MAE) of -13.2 meV/f.u.. The $Mn_2Ge_2Te_6$ shows good dynamical and thermal stability. Moreover, $Mn_2Ge_2Te_6$ presents good ferromagnetic and half-metallic stability under charge doping. The carriers doping could effectively tune magnetic and electronic properties. Specifically, the magnetic moment, exchange parameter, and MAE could be efficiently tuned. The total magnetic moment changes linearly with charges doping. The exchange parameters could be controlled by the doping carriers. The carriers doping could modulate MAE to -18.4 (+0.4 $e$), -0.85 (-1.6 $e$), 1.31 (-2.4 $e$) meV/f.u., by changing hybridization between Te atom's $p_y$ and $p_z$ orbitals. $Mn_2Ge_2Te_6$ with intrinsic ferromagnetism, high tunable MAE, good stability of ferromagnetism and half-metallicity could help researchers to investigate its wide application in the electronics and spintronics.

## 1. INTRODUCTION

Two-dimensional (2D) intrinsic ferromagnetic materials, especial HM is important for the spintronics.[1-3] 2D materials, such as Graphene,[4] h-BN,[5] $MoS_2$,[6, 7] and Mxenes[8] have been successfully synthesized in the experiments. However, 2D ferromagnetic materials are rare.[9, 10] This situation is limited by the Mermin-Wagner theory,[11] which says 2D magnetic materials cannot exist in the isotropic Heisenberg model at finite temperature. Until recent years, $FePS_3$,[12] $CrI_3$,[13, 14] $VSe_2$,[15] $FeGeTe_2$,[16, 17] $CrSe_2$[18] and $CrGeTe_3$ (CGT)[19, 20] monolayer (ML) with intrinsic ferromagnetism have been successfully synthesized in the experiments. 2D magnetic materials are becoming research hot, as the 2D magnetic materials have far-reaching implications for the basic physics and engineering such as spintronic devices.[21] 2D magnetic materials are expected to have amusing properties,[22] such as high Curie ($T_c$),[23] large magnetic crystalline anisotropy energy (MAE), high spin polarization and controllable electromagnetic properties.[24] For Half-metallic materials,[25] one spin channel is insulating or semiconducting, while another channel is conducting.[25] Therefore, half-metallic materials could get 100% spin-polarized current, which are high desired in the spintronics. HM could be work as pure spin generated and injected devices. The idea HM used in the spintronics is expected a high $T_c$, and the semiconductive gap should be large enough to prevent the thermally agitated spin-flip transition and preserve half-metallicity at room temperature.[22] Furthermore, the large MAE is urgently needed for the magnetoelectronics.[26]

Most 2D materials are semiconductors, or common metals. Low-dimensional materials, such as graphene nanoribbon (GNR), could be transformed into HM with an external electric field.[27] The chemically functioned GNR could be transformed into HM,[28] but these supposes are hard to realize in the experiments.[29, 30] Among synthesized 2D magnetic materials, $CrI_3$ is a semiconductor with $T_c$ of 45 K,[31] while $VSe_2$,[18] $Fe_3GeTe_2$,[17] $CrSe_2$[18] and CGT[19] are normal spin-polarized metal with FM order. Furthermore, the electronic properties of $VSe_2$ are dominated by the substrates.[32] The $FePS_3$ shows antiferromagnetic order.[33] In sum, HM is quite rare in 2D materials.[9] However, 2D HM is highly desired in the spintronics.[34, 35] So we construct and study half-metallic $Mn_2Ge_2Te_6$ ML with intrinsic ferromagnetism by density functional theory and a global minimum search.

In this work, the electronic and magnetic properties of $Mn_2Ge_2Te_6$ are systematically investigated by using first-principles method. $Mn_2Ge_2Te_6$ ML is an intrinsic ferromagnetic material owing to the superexchange interaction between Mn and Te atoms, with $T_c$ of 316 K. It can be concluded by the Goodenough-Kanamori-Anderson (GKA) theory. $Mn_2Ge_2Te_6$ is HM with a band gap of 1.462 eV in spin-β channel. $Mn_2Ge_2Te_6$ shows IPA, with MAE of -13.19 meV/f.u. Besides that, $Mn_2Ge_2Te_6$ ML retains half-metallicity with FM order under charges doping. More strikingly, magnetic moment, energy difference between different magnetic orders, exchange parameters, and MAE of $Mn_2Ge_2Te_6$ could be easily modulated by the charges doping. The magnetic moment of $Mn_2Ge_2Te_6$ changes linearly with carriers doping. The $Mn_2Ge_2Te_6$ ML still shows FM order, and the ferrimagnetic-stripy order has the second lowest energy in the considering orders. The magnetic exchange parameters including $J_1$, $J_2$, $J_3$, represent the first, second, and third in-plane nearest-neighbor spin-spin exchange interactions, respectively. $J_1$, $J_2$, $J_3$ are (0.005, 0.002, 0.005), (0.017, 0.011, 0.013), and (0.028, 0.007, 0.012) eV, with -1.8, 0.6, and 1.5 $e$ charges doping, respectively. The MAEs monotonously decrease due to the energy reduce of Te's $p_y$ and $p_z$ orbitals hybridization with electrons doping. The EA could be eventually switched from in-plane to out-of-plane ($q$>-2.056 $e$). The MAEs increase to the maximum of -18.42 meV ($q$ < +0.4 $e$) at first, and then decrease as the hybridization between Te's orbitals changes with more holes doping in $Mn_2Ge_2Te_6$.

## 2. COMPUTATIONAL DETAILS

The calculation of $Mn_2Ge_2Te_6$ are using plane-wave basis Vienna Ab initio Simulation Package (VASP) code,[36] based on the density functional theory. The generalized gradient approximation (GGA) with Perdew-Burke-Ernzerhof (PBE)[37] is adopted. Mn 3$d$ electron is dealt with hybrid-functional HSE06[38, 39] and GGA+U method.[40] The energies with different orders, band structures, density of states (DOS), and $T_c$ are calculated by HSE06 functional. And MAE, phonon spectra, and molecular dynamics are performed by LDA+U method. The effective onsite Coulomb interaction parameter ($U$) and exchange interaction parameter ($J$) are set to be 4.60 and 0.60 eV, respectively. Therefore, the effective $U_{eff}$ ($U_{eff} = U - J$) is 4.00 eV.[41, 42] And the corresponding magnetic orders and electronic properties are consistent with HSE06 functional, shown in Figure S1. The vacuum space in the $z$-direction is set 16 Å to avoid the virtual interactions. The kinetic energy cutoff is set as 300 eV for optimizing geometry and calculating energy. The geometries are fully relaxed until energy and force is converged to $10^{-6}$ eV and 1 meV/Å, respectively. 9×9×1 and 16×16×1 Monkhorst-Pack grids[43] are used for geometry optimization and energy calculation, respectively. The MCA energy is calculated with an energy cutoff of 400 eV and convergence of $1\times10^{-8}$ eV for the total energy. The spin-orbital coupling (SOC) is also considered in MCA calculation, and the corresponding $k$-grid is adopted 19×19×1, without any symmetry constriction. The $k$-grid is systematically tested, shown in Figure S2. The phonon spectra and DOS are calculated using finite displacement method as implemented in the Phonopy software.[44] A 4×4×1 cell is adopted in the calculation. The total energy and Hellmann-Feynman force is converged to $10^{-8}$ eV and 1 meV/Å in the phonon spectra calculation, respectively. 6000 uniform $k$-points along high-symmetry lines are used to obtain phonon spectra. *Ab initio* molecular dynamics (AIMD) simulation is also performed to confirm structural dynamical stability. The constant moles–volume–temperature (NVT) ensemble with Nosé–Hoover thermostat[45] is adopted at temperature of 300 and 500 K, respectively. The time step and total time is 1 fs and 10 ps, respectively. In order to eliminate the effect of the periodic boundary condition with relatively smaller system size, which can artificially increase the stability of the structures, a larger supercell (2×2×1 cell) is used in the AIMD simulation.

## 3. RESULTS AND DISCUSSION

**3.1. Geometry of $Mn_2Ge_2Te_6$ ML.** The geometry of $Mn_2Ge_2Te_6$ ML is fabricated, and confirmed by particle swarm optimization (PSO)[46] based on the crystal structure analysis, and optimized structure is provided in Figure 1 a-c. The corresponding optimized lattice parameter is $a = b = 6.968$ Å, by fitting energy with lattice parameters, which is larger than 5.989 Å of CGT.[19, 47] And more detail could be found in Figure S3. The bond length between Mn and Te atoms is 2.915 Å, while the bond length between Ge and Te atoms is 2.617 Å. The bond length between Ge and Ge atoms is 2.477 Å. From the optimized geometry, we can find that $Mn_2Ge_2Te_6$ ML presents $D_{3d}$ point group, which is the same with CGT.[19]

The Mn atom is in the center of the octahedron, like Cr atom in CGT. There is 1.033 $e$ electron transfer from Mn atom to Ge (0.409 $e$) and Te (0.695 $e$) atoms by the bader analysis.[48] The Te atoms get more electrons, as Te atom shows more stronger electronegativity than Ge atom. The Mn atom shows $3d^64s^1$ configuration. When one $d$ electron is taken away, it results in $Mn^{1+}$ ions. Mn atom has a high-spin octahedral $d^6$ configuration, leading to a large magnetic moment of 4.365 $\mu_B$, while Ge and Te atoms have -0.020 (0.01×2) and -0.876 (0.146×6) $\mu_B$, respectively. The corresponding spin charge density difference is shown in Figure 1 d-e, respectively. In a word, the magnetic moment mainly localizes in Mn atoms, shown in Figure 1 d-e. Each supercell has two Mn atoms. Therefore, there are two kinds of magnetic orders: FM and antiferromagnetic (AFM) orders. The total magnetic moment is 8.00 $\mu_B$ for FM order, while the total magnetic moment is 0.00 $\mu_B$ for AFM order. And the corresponding spin charge density difference with FM and AFM orders are shown in Figure 1 d-e, respectively. In order to describe magnetic stability, the energy difference ($\Delta E$) between FM and AFM orders is defined: $\Delta E = E_{FM} - E_{AFM}$. The $\Delta E$ is 0.123 eV, which implying $Mn_2Ge_2Te_6$ has FM ground state.

Why does $Mn_2Ge_2Te_6$ show FM order? Each Mn atom is coordinated by six ligands-Te in $Mn_2Ge_2Te_6$. The corresponding Te-Mn-Te bond angle is 92.72°, 82.71°, 102.37°, and it causes FM coupling (shown in Figure 1f, g), according to the Goodenough-Kanamori-Anderson rules[49-51] of superexchange theorem. However, there is direct exchange interaction between Mn and nearby Mn atoms, which favors AFM coupling, shown in

Figure 1f. The ground state is determined by the competition between superexchange and direct exchange interaction, similar to $CrI_3$[31] and CGT. The superexchange interaction is stronger than the direct exchange interaction in $Mn_2Ge_2T_6$. In other words, the superexchange interaction originating from the hybridization between Mn-$d$ and Te-$p$ orbitals dominates the exchange interaction, shown in Figure S4 a-d. Finally, $Mn_2Ge_2Te_6$ shows FM ground state.

The geometry and magnetic properties of $Mn_2Ge_2Te_6$ are investigated in the above section, but the electronic properties are usually related with the geometry. The band structure and partial density of the states (PDOS) of $Mn_2Ge_2T_6$ are calculated, shown in Figure 1 h-i. The spin-α electron channel is conducting, while the spin-β electron channel is insulating. Therefore, the $Mn_2Ge_2Te_6$ is HM. The spin-α electrons partially occupy the Fermi-level. For the spin-β electrons, the valance band maximum locates at Γ point, while the conductance band minimum locates at K point. Therefore, the $Mn_2Ge_2Te_6$ is a semiconductor with an indirect gap of 1.462 eV for the spin-β electrons, shown in Figure 1h. The larger band gap of one spin channel could effectively prevent spin-leakage.[52] Furthermore, 100% spin-polarization implies $Mn_2Ge_2Te_6$ could work as spin injection and spin transport devices.[53] Through further analysis, the states near the Fermi-level are mainly contributed by the Te's $p$ orbitals, while the states near the Fermi-level are partially contributed by Mn's $d_{xy}$, $d_{yz}$, $d_{x^2-y^2}$ and $d_{xz}$ orbitals, shown in Figure 1i, S4, S5, respectively. The PDOS and integrated density of the states (IDOS) of Mn atoms are shown in Figure S5a, b, respectively. From Figure S4, S5, it can be found that spin-α electrons are conductive, while the spin-β electrons are insulating, which is consistent with the above analysis.

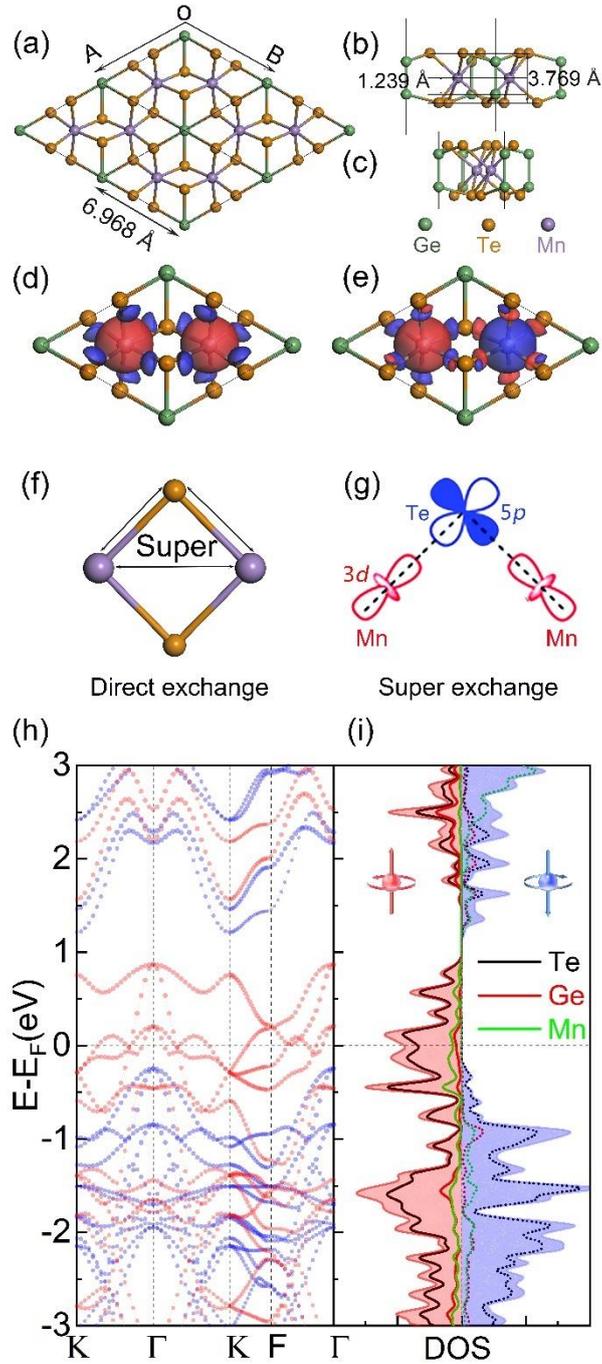

**Figure 1.** (a) Top, (b) side-1 (along *x* axis) and (c) side-2 (along *y* axis) views of optimized geometries of $Mn_2Ge_2Te_6$ ML. The green, yellow and blue balls represent Ge, Te, and Mn atoms, respectively. (d-e) Spin charge densities difference of (c) FM and (d) AFM orders of $Mn_2Ge_2Te_6$ ML. The isovalue is 0.02 e/Å³. (f) Direct and (g) superexchange interaction. (h) The spin-polarized band structures, and the red and blue lines represent spin-α and spin-β electrons. (i) PDOS with FM order. The black, red, and green lines represent partial density of the Te, Ge, and Mn atoms, respectively.

**3.2. Magnetic and Electronic Properties.** In the application of 2D materials, magnetic and electronic properties are important. The different magnetic configurations are investigated to ascertain the magnetic ground state, shown in Figure 2a-d. Each Mn atom in $Mn_2Ge_2Te_6$ contributes 4.0 $\mu_B$ magnetic moment. There are eight Mn atoms in the 2×2×1 cell. Therefore, there is 32.0 $\mu_B$ magnetic moment for the FM order. Three different AFM orders are considered, including AFM-zigzag (AFM-Z), AFM-stripy (AFM-S), and AFM-Néel (AFM-N) orders. For the AFM orders, four Mn atoms contribute 16.0 $\mu_B$ magnetic moment, while the other four Mn atoms contribute -16.0 $\mu_B$ magnetic moment. And the magnetic moment shows different distribution. As a result, the total magnetic moment equals to 0.0 $\mu_B$. The spin charge density difference is shown in Figure 2 a-d. And energy difference is defined as the energy difference between AFM and FM orders. The highest energy with AFM-Z order is 0.637 eV higher than FM order, and AFM-N order has the second highest energy of 0.614 eV, shown in Figure 2 b, d, respectively. The AFM-S order is 0.457 eV higher than FM order, which has the lowest energy in the AFM orders, shown in Figure 2c.

The $T_c$ of ferromagnetic materials is calculated using classic Heisenberg model Monte Carlo (MC) with the following formulas:

$$H = -J \sum_{<i,j>} S_i * S_j \quad (1)$$

$$E_{FM} = E_0 - (3J_1 + 6J_2 + 3J_3)|S|^2 \quad (2)$$

$$E_{AFM\text{-}Néel} = E_0 - (-3J_1 + 6J_2 - 3J_3)|S|^2 \quad (3)$$

$$E_{AFM\text{-}zigzag} = E_0 - (J_1 - 2J_2 - 3J_3)|S|^2 \quad (4)$$

$$E_{AFM\text{-}stripy} = E_0 - (-J_1 - 2J_2 + 3J_3)|S|^2 \quad (5)$$

Where $E_{FM}$, $E_{AFM\text{-}Néel}$, $E_{AFM\text{-}zigzag}$, and $E_{AFM\text{-}stripy}$ represent energies with FM and AFM-N, AFM-Z and AFM-S orders, respectively. $J$ and $H$ are the exchange parameter and Hamilton, respectively, and $S_i$ presents the spin operator, shown in Figure 2e. For $Mn_2Ge_2Te_6$ ML, the corresponding $J_1$, $J_2$ and $J_3$ is 13.6, 7.5, 12.0 meV, respectively. Both nearest- and next nearest-neighbor Mn atoms show FM couplings. However, $J_1$, $J_2$ and $J_3$ of CGT is 2.71, -0.058, and 0.115 meV,[19] respectively. And the corresponding MC code is developed by Prof. Hongjun Xiang' group.[54] As benchmark, the $T_c$ of CrI$_3$[31] is calculated to be 51 K with this method, which agrees well with the experimental result of 45 K. A larger 80×80 cell with 1.0×10$^8$ loops is used to evaluate $T_c$. The 4.0 $\mu_B$ magnetic moment per Mn atom drops quickly. And the corresponding $T_c$ are predicted to be 316 K, which is higher than CGT (bulk, 66 K).[19]

The electronic properties are usually related with the magnetic orders. The FM order is HM, while all AFM orders are spin-unpolarized metal, shown in Figure S6 a-d. Though Mn$_2$Ge$_2$Te$_6$ with different AFM orders are spin-unpolarized metal, they are different from each other. More discussion could be found in the Supporting Information.

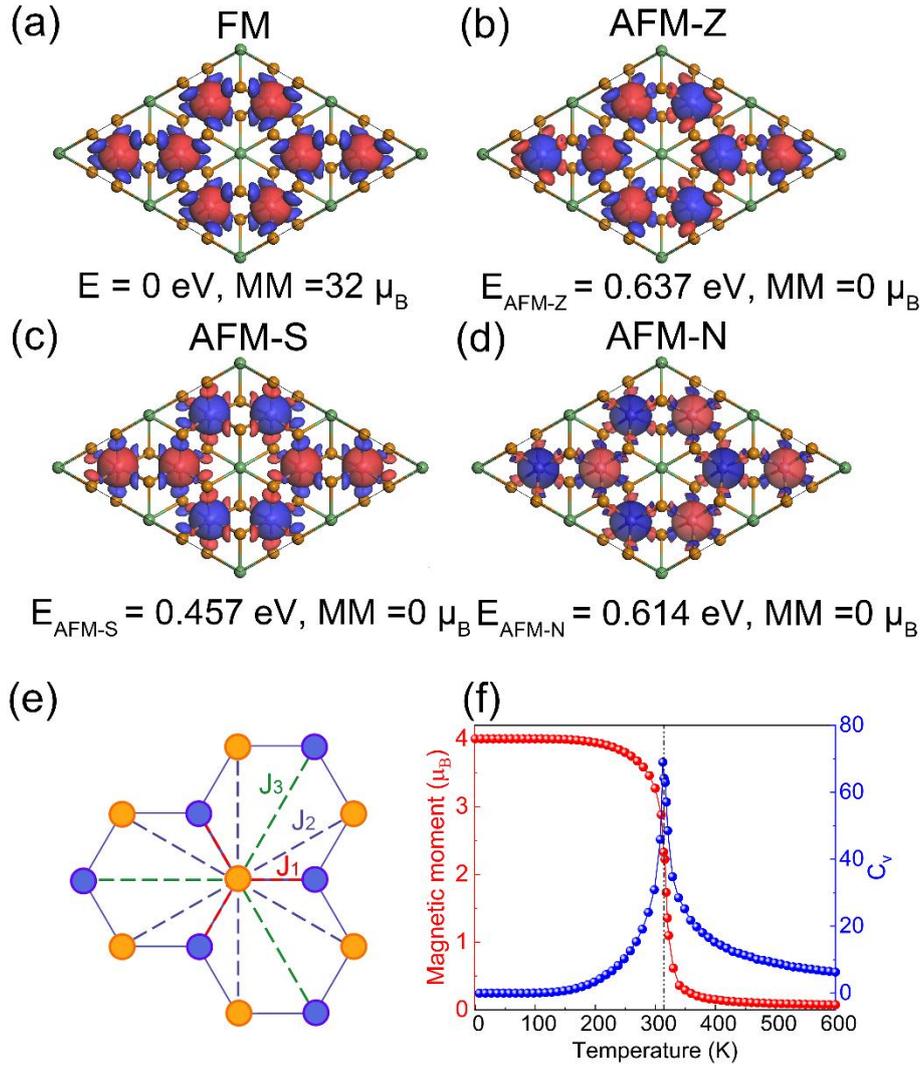

**Figure 2.** The spin charge density difference of $Mn_2Ge_2Te_6$ with (a) FM, and (b) AFM-Z, (c) AFM-S, and (d) AFM-N orders. The red and blue represent spin-α and spin-β electrons, respectively. (e) Crystal structure consisting of magnetic ion Mn only. Illustration of neighbor exchange interactions. $J_1$, $J_2$ and $J_3$ represent the first, second, and third in-plane nearest-neighbor spin-spin exchange interactions, respectively. (f) Magnetic moment per unit cell (red) and specific heat ($C_v$) (blue) vary respect to the temperature from Heisenberg model MC simulation, respectively.

### 3.3. Magnetic Anisotropy Properties.

The idea HM is expected to have higher $T_c$ and larger MAE.[22] In this part, the MAE and magnetocrystalline anisotropy (MCA) are calculated using LDA+U method. MCA is MAE per area. The adopted magnetic materials in the spintronics are expected to have higher MCA, which means electron needs more energy to overcome a higher "barrier" from EA to hard axis.[35] In a word, MCA is benefit for preserving the direction of magnetic moments from heat fluctuation. As $Mn_2Ge_2Te_6$ has $D_{3d}$ point group, the corresponding energy ($E$) along certain direction ($\theta$, $\phi$) follows the following equation:

$$\Delta E_0 = K_1 \cos^2\theta + K_2 \cos^4\theta + K_3 \cos 3\phi \quad (6)$$
$$\Delta E_0 = E - E_{[001]} \quad (7)$$

where $E_{[100]}$ represents the energy along [100] direction. $K_1$ and $K_2$ show the contribution of the quadratic and quartic part to MAE, respectively. The energy difference $\Delta E_0$ is independent of the in-plane azimuthal angel $\phi$, so $K_3$ equals to 0, shown in Figure 3 a-b. The eq 6 is simplified into the following equation: [55]

$$\Delta E_0 = K_1 \cos^2\theta + K_2 \cos^4\theta \quad (8)$$

The $\Delta E_0$ changes as a function of polar angle $\theta$, shown in Figure 3c. And $\Delta E_0$ follows the equation: $\Delta E_0 \text{ (meV)} = -12.56\cos^2\theta + 0.492\cos^4\theta$ for $Mn_2Ge_2Te_6$ ML. Therefore, the MAE and MCA could be calculated using followed equations:

$$MAE = E_{[100]} - E_{[001]} \quad (9)$$
$$MCA = E_{[100]} - E_{[001]} = MAE / S \quad (10)$$

$E_{[001]}$ represent the energy with magnetic axis along [001] direction. $S$ is the area of the supercell. $S$ is calculated in this equation: $S = a^2 \sin 60°$, and $a$ is lattice parameter of unit cell. The corresponding MAE and MCA is -13.2 meV and -5.029 erg/cm$^2$, respectively. And the negative MAE implies EA of Mn$_2$Ge$_2$Te$_6$ points to in-plane direction, shown in Figure 3d. Compared with CGT (MAE = 0.5 meV),[56] the magnetic anisotropy of Mn$_2$Ge$_2$Te$_6$ is obviously reinforced, which origins Mn atom (54.938) is heavier than Cr atom (51.996). Therefore, the corresponding SOC (including direct and indirect SOC) is stronger. MAE and MCA mainly come from the contribution of indirect SOC, similar with CrI$_3$, and VSeTe.[26]

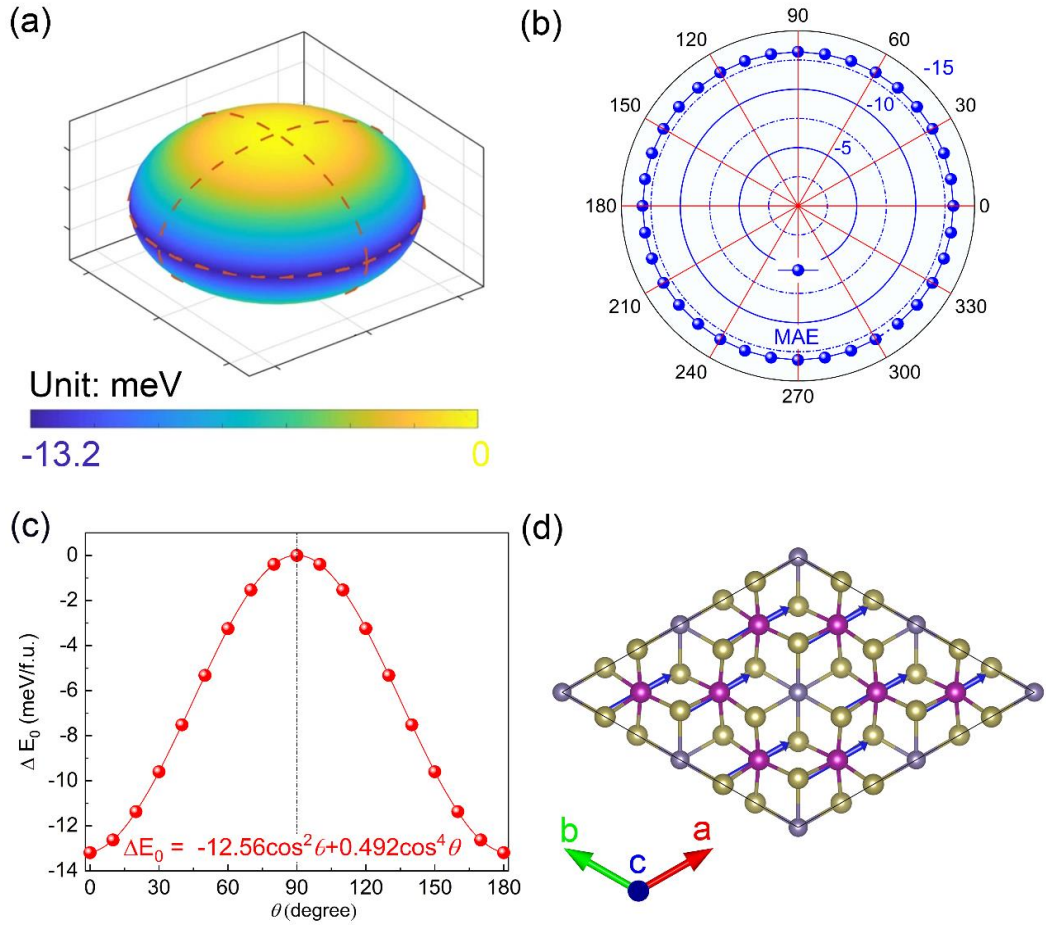

**Figure 3.** The MAE map ([001] FM state as a reference) of $Mn_2Ge_2Te_6$. (a) $\Delta E_0$ varies from the out-of-plane to the in-plane direction. (b) The energy indicated by the dashed lines changes with azimuthal angle $\varphi$. (c) The $\Delta E_0$ changes with polar angle $\theta$. (d) The blue arrow represents direction of EA (along [100] direction).

**3.6. The Dynamical and Thermal Stability.** The dynamical stability of $Mn_2Ge_2Te_6$ is confirmed via phonon dispersion curves and phonon DOS, which show no obvious imaginary phonon modes. The highest vibration frequency is 7.388 THZ, which is lower than CGT (8.364 THZ), shown in Figure 5a, S7. From Figure 5b, we can find that the contribution mainly comes from Te atoms for the low frequency ($0<\varepsilon<3$ THZ). On the contrary, Ge atoms make much contribution to the high frequency ($6<\varepsilon<8$ THZ) parts. Mn atoms mainly contribute at the middle frequency ($4<\varepsilon<5$ THZ).

The thermal stability of $Mn_2Ge_2Te_6$ is also evaluated with AIMD. To examine the stability of geometry and magnetic order at room temperature, we also perform AIMD simulation at 300 and 500 K, respectively. The fluctuation in the total energies is also evaluated. The total energies vibrate round -173.05 eV at 300 K, and -171.73 eV at 500 K, with the amplitude about 0.025 and 0.052 eV per atom, shown in Figure 5c-f. And the snapshots of the geometries also confirm the essential intact structures, shown in Figure S8. No obvious structure destruction is found, so $Mn_2Ge_2Te_6$ should be stable at 300 K. In addition, the evaluation of magnetic moment changes with the temperature is also investigated, and the total magnetic moment is about 33.0 $\mu_B$ in the simulation, shown in Figure 4 d, f. Therefore, $Mn_2Ge_2Te_6$ ML presents ferromagnetic ground, and geometry is stable at room temperature or higher temperature (500 K).

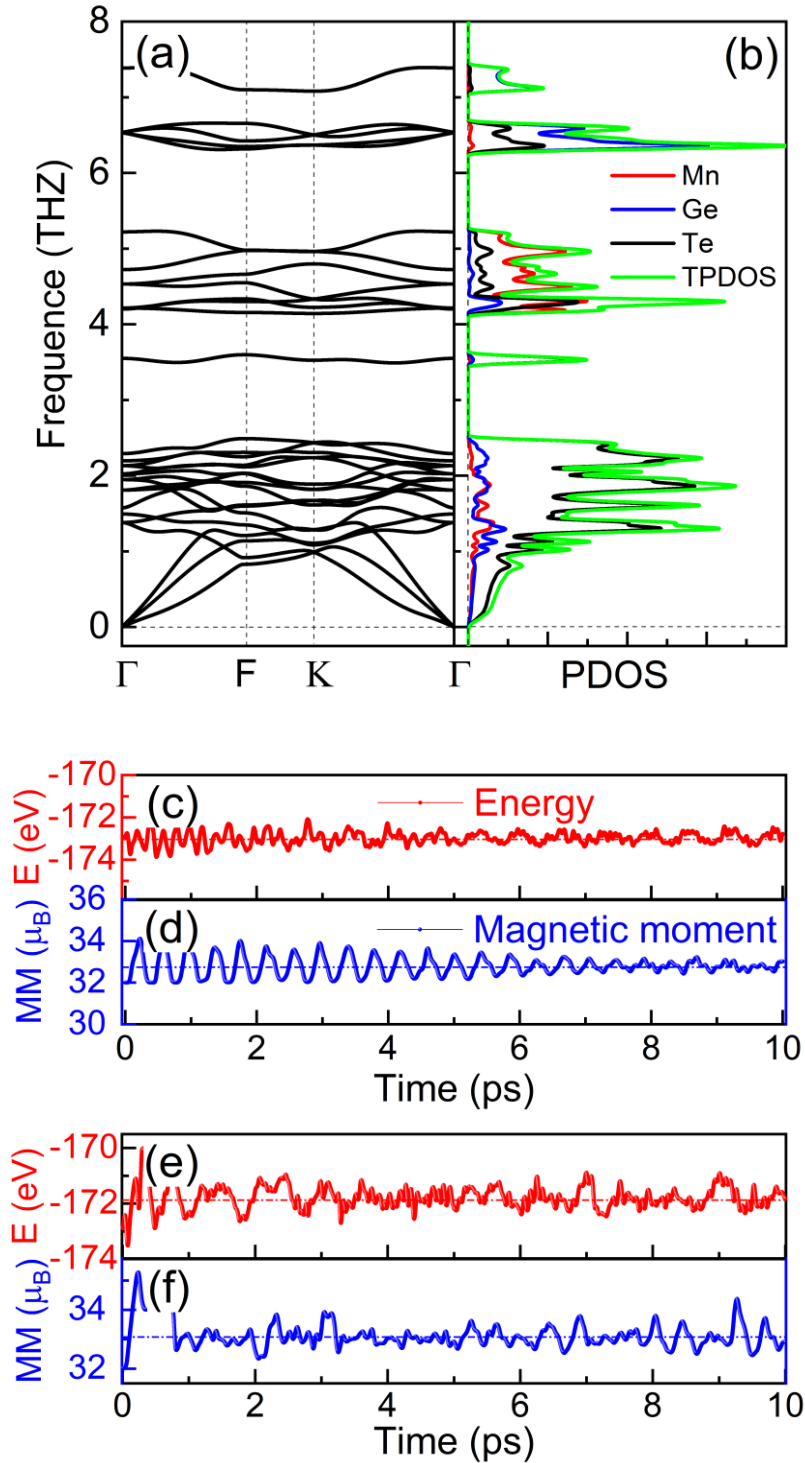

**Figure 4**. (a) The phonon band structure and (b) density of the states of Mn$_2$Ge$_2$Te$_6$. (c) The total energy (red color) and (d) magnetic moment change (blue color) with the times at simulated 300 K. (e) The total energy and (f) magnetic moment change with the time at 500 K.

**3.5. Magnetization and Curie Temperature Modulation.** The carrier doping could effectively tune magnetic properties of 2D materials.[35, 56] The magnetic moment of $Mn_2Ge_2Te_6$ could be effectively controlled by the electrons and holes. The total magnetic moment could be evaluated as: MM = $9-|q|$, shown in Figure 5a. The $q$ is the doped charges. The similar trend also appears in other low-dimensional materials.[35] When $q$ = -1.0, -0.6, -0.2, 0.6, 1.0, the corresponding total magnetic moment is 9.0, 8.6, 8.2, 7.4, and 7.0 $\mu_B$, respectively. And the magnetic moment mainly localizes in Mn atoms, shown in Figure 5 b-f. When -1.0 (spin charge density difference is shown in Figure 5b), -0.6 (Figure 5c), -0.2, +0.6 (Figure 5d) and +1.0 $e$ (Figure 5e) charges are doped in $Mn_2Ge_2Te_6$, each Mn atom has 4.434, 4.406, 4.379, 4.328, 4.297 $\mu_B$, show in Figure 5a. While each Ge and Te atoms have 0.026 (-0.061), 0.022 (-0.099), 0.014 (-0.131), 0.008 (-0.195), and 0.015 (-0.229) $\mu_B$, respectively. And the corresponding spin charge density difference is shown in Figure 5 b-f. As more electrons are doped, the Te atom gets more electrons, shown in Figure S10. Therefore, the corresponding Te's magnetic moment increases, shown in Figure 5a. However, Mn atoms' magnetic moment monotonously decreases. It origins Te's electrons are taken away, and more detail is shown in Figure S10.

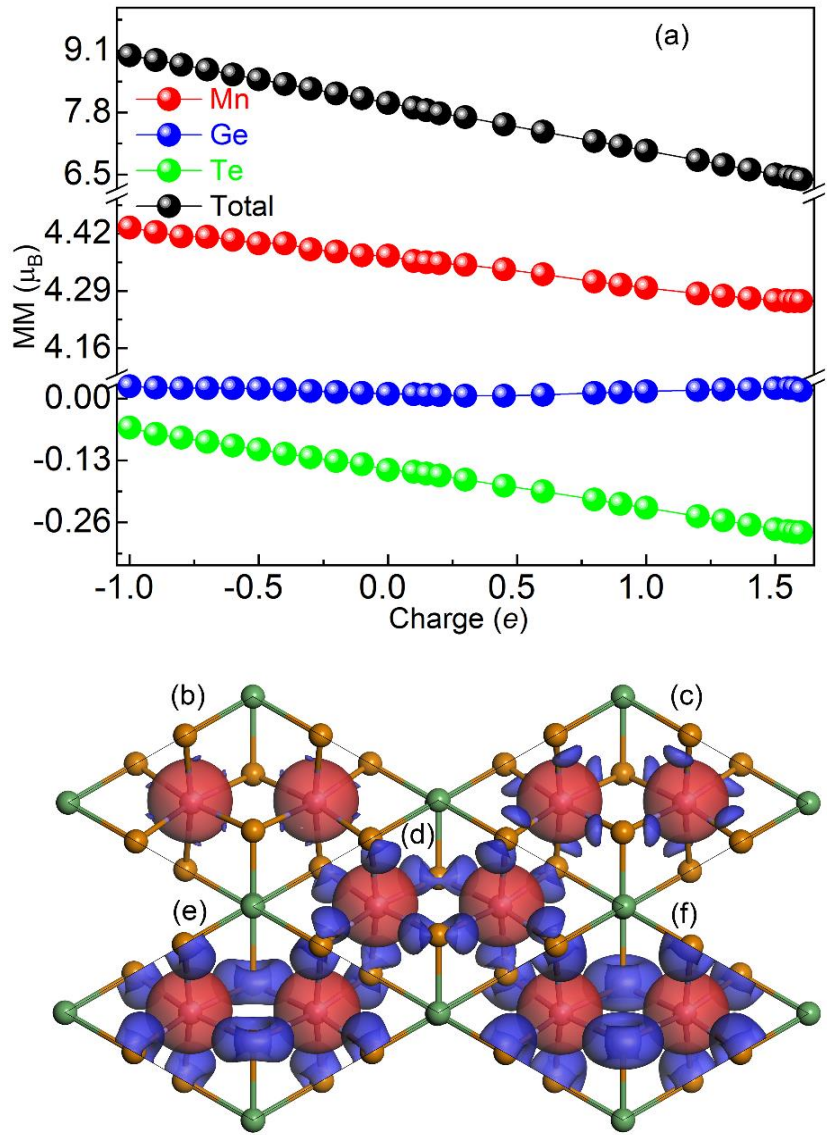

**Figure 5.** (a) The evolution of magnetic moment of Mn, Ge, and Te atoms of Mn$_2$Ge$_2$Te$_6$. The red, blue, green and black lines with dots represent magnetic moment of Mn, Ge, Te and total atoms, respectively. The spin charge density difference of Mn$_2$Ge$_2$Te$_6$ with charges doping of (b) -1.0, (c) -0.6, (d) 0.6, (e) 1.0 and (f) 1.55 *e*, respectively. The red, and blue represent spin-α and spin-β electrons, respectively. The isovalue is set 0.03 e/Å$^3$.

In $Mn_2Ge_2Te_6$, the FM order could be enhanced or weakened by the hole/electron doping, and similar trend also appears in $CrSe_2$.[57] And the corresponding energies with different magnetic orders also change. Original AFM orders change into Ferrimagnetic (Ferrim) orders. The energy difference between all kinds of Ferrim and FM orders change with charges doping, shown in Figure 6a. $J_1$, $J_2$, and $J_3$ also change with charges doping, shown in Figure 6b, respectively. $Mn_2Ge_2Te_6$ maintains FM ground state for the carriers doping, while Ferrim-Stripy still has the second lowest energy. The smallest energy difference between FM and Ferrim-Stripy orders ($\Delta E_{Ferrim-S} = E_{FM} - E_{Ferrim-S}$) also changes with the charges doping. When -0.5, -1.0, -1.8 $e$ electrons are injected into $Mn_2Ge_2Te_6$ ML, the corresponding $\Delta E_{Ferrim-S}$ is 0.521, 0.386, and 0.132 eV, respectively. The corresponding energy difference between FM and Ferrim-Neel ($\Delta E_{Ferrim-N} = E_{FM} - E_{Ferrim-N}$), Ferrim-Zigzag ($\Delta E_{Ferrim-Z} = E_{FM} - \Delta E_{Ferrim-Z}$) orders is 0.758 (0.662), 0.708 (0.598), and 0.228 (0.211) eV, respectively. The corresponding $J_1$, $J_2$, and $J_3$ is (19.3, 6.6, 12.3), (15.5, 4.3, 14.0), (4.6, 1.8, 4.9) meV, respectively. As 0.40, 0.80, 1.20, and 1.50 $e$ electrons are "pumped away", the corresponding $\Delta E_{Ferrim-S}$ is 0.563, 0.653, 0.694, and 0.675 eV, respectively. The corresponding $\Delta E_{Ferrim-N}$ and $\Delta E_{Ferrim-Z}$ is 0.706 (0.715), 0.774 (0.815), 0.930 (0.840) and 0.967 (0.749) eV, respectively. And $J_1$, $J_2$, and $J_3$ is (17.3, 8.9, 12.1), (19.1, 10.8, 13.1), (24.5, 9.4, 14.3) and (27.9, 7.1, 12.4) meV, respectively. This origins that the carrier doping could effectively control the superexchange interaction by changing the *p-d* hybridization between Mn and Te atoms.[47]

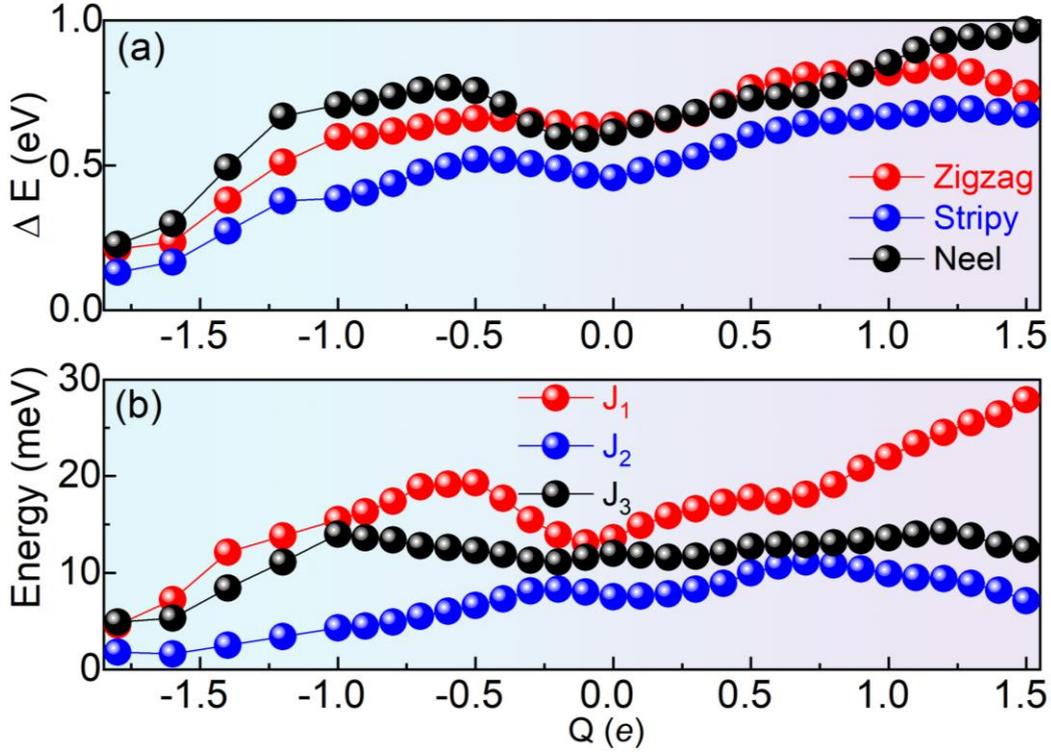

Figure 6. The energy difference between Ferrim and FM orders and $J_1$, $J_2$, $J_3$ change with the changes doping, (a) The $\Delta E_{\text{Ferrim-N}}$, $\Delta E_{\text{Ferrim-Z}}$ and $\Delta E_{\text{Ferrim-S}}$ change with the charges doping. (b) The $J_1$, $J_2$, $J_3$ change with changes doping. The red, blue and black lines with dots represent $J_1$, $J_2$, and $J_3$, respectively.

The electronic properties are usually related with the magnetic orders, and $Mn_2Ge_2Te_6$ shows robust half-metallicity under charge doping, shown in Figure 7. The spin-β electrons of $Mn_2Ge_2Te_6$ ML are still semiconductive, while the spin-α electrons are always conductive, shown in Figure 7 a-f. $Mn_2Ge_2Te_6$ doped by charges is always HM, while only the states at the Fermi-level change with charges doping, which implying conductivity may change. The DOS doped with -1.0, -0.6, -0.3, 0.3, 0.6, 1.0 $e$ charges are also calculated, shown in Figure 7 a-f, respectively. All spin-α electron channel is conducting, while spin-β electron is insulating. Therefore, they are all half-metal (HM), as there is a large gap, for spin-β electrons channel. However, the DOS of spin-α electrons at Fermi-level is different from each other. When -0.3 and -0.6 $e$ electrons are doped, the corresponding Fermi-level is moved upward? And the corresponding DOS at Fermi-level is decreased to 1.842 arb. unit, shown in Figure 7 b, c. As -1.0 $e$ electron is doped, the corresponding DOS at Fermi-level is further decreased to 0.783 arb. unit, shown in Figure 7a. For the holes doping of 0.3, and 0.6 $e$, it means that electrons are "pumped away". As a result, the Fermi-level is shifted downward. And the DOS at Fermi-level is further increased to 2.137, and 2.163 arb. unit, shown in Figure 7d, e. The different DOS at Fermi-level implies different conductivity under different charges doping. At last, the $Mn_2Ge_2Te_6$ shows high good half-metallic stability under charge doping, while the conductivity changes.

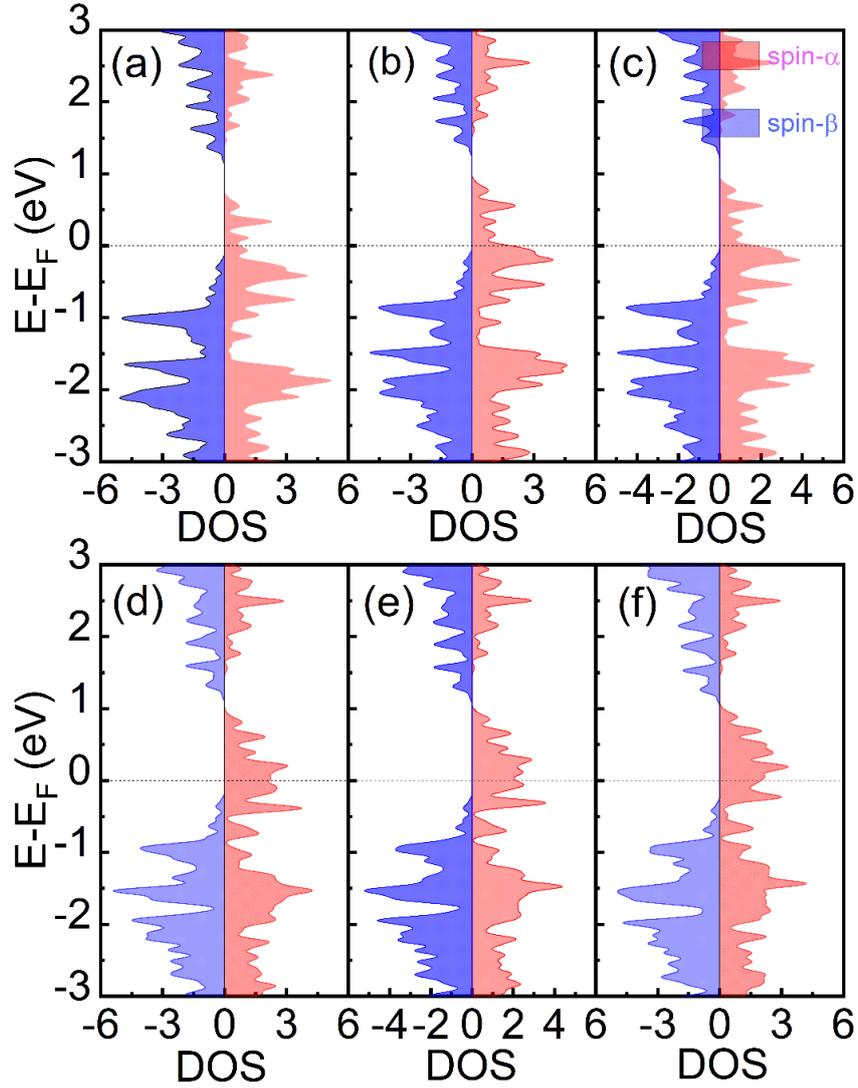

**Figure 7**. The DOS of $Mn_2Ge_2Te_6$ is doped with (a) -1.0, (b) -0.6, (c) -0.3, (d) +0.3, (e) +0.6, and (f) +1.0 *e* carriers, respectively. The red and blue colors represent spin-α and spin-β electrons, respectively. The Fermi-level is set 0 eV.

**3.6. Magnetocrystalline Anisotropy Modulation.** Generally speaking, when the 2D materials are synthesized, they are usually doped by the carriers. In addition, the carrier doping is widely used to modulate magnetic anisotropy properties of 2D materials.[58] The DOS near the Fermi-level are mainly dominated by the Te's *p* orbitals. Therefore, the charge redistribution by the carriers doping could affect DOS near the Fermi-level. Thus, the MAE and MCA could be controlled by the injected charges in 2D materials, such as CGT,[56] and $CrSe_2$.[18] The EA of $Mn_2Ge_2Te_6$ could be rotated from [100] to [001] direction, as much electrons are injected into $Mn_2Ge_2Te_6$ ML, shown in Figure 8 a-b, and Figure S11 a-i. However, MAE could be boosted as more holes doping ($q < 0.4\,e$), shown in Figure 8 a-e. No matter electrons or holes dope $Mn_2Ge_2Te_6$, the eq 7 still works, shown in Figure 8 b-e. When 0.2, 0.5, 1.0, 1.6, and 2.0 *e* electrons are injected, the corresponding MAEs are -11.47, -10.29 (Figure 8c), -3.68, -0.852 and -0.161 meV, respectively. As -2.055 *e* electrons are injected, the MAEs are further decreased to -0.043 meV, shown in Figure S11d. When -2.056 *e* electron is injected, the corresponding MAE is 0.002 meV, which means that EA is switched from the in-plane to the out-of-plane, shown in Figure S11e. MAE is important for the magnetic information storage, and perpendicular magnetic anisotropy (PMA) is useful for the magnetic information storage with high density.[59, 60] Most interestingly, the MAEs follow this equation: [55]

$$\Delta E_0 = K_1 \cos^2\theta + K_2 \cos^4\theta + K_3 \cos^6\theta \qquad (11)$$

when the MAE is near the 0 meV. Taking q = -2.055, −2.056 $e$ as an example, the corresponding $\Delta E_0$ follows these equations: $\Delta E_0 = 4.651\cos^2\theta - 6.404\cos^4\theta + 1.790\cos^6\theta$ ($10^{-1}$ meV), $\Delta E_0 = 4.578\cos^2\theta - 6.073\cos^4\theta + 1.553\cos^6\theta$ ($10^{-1}$ meV), shown in Figure S11 d, e, respectively. And more detail could be found in Figure S11, in the Supporting Information. In the doping range of [-2.10, -2.00] $e$, the MAE and doping charges $q$ displays a linear change:

$$\text{MAE} = -6.2 - 3.02q \text{ (meV)} \qquad (12)$$

And the corresponding MAEs change with charges doping, shown in the inset of Figure 8a and Figure S11 j-l. As more negative carriers, such as 2.2 (Figure S11k), and 2.3 $e$ electrons are further doped, the MAEs are further increased to 0.485, and 0.850 meV, respectively. When 2.4 $e$ electron is injected, the corresponding MAE is 1.31 meV, shown in Figure 8a. More detail could be found in Figure S11.

For the positive carriers (holes) doping, the corresponding MAEs are firstly enhanced, and then weakened, shown in Figure 8 a, d, e, respectively. The MAEs are -14.2 (+0.075 $e$), -16.62 (+0.20 $e$), -18.0 (+0.30 $e$), respectively. As +0.4 $e$ hole is doped, the corresponding MAE reaches the largest value (-18.42 meV), shown in Figure 8d. When more positive carriers are doped, the corresponding MAEs decrease again. As 0.6, 1.0, 1.6 *e* electrons are taken away, the corresponding MAEs are decreased to 17.41, 12.77 (Figure 8e), and 8.83 meV, respectively.

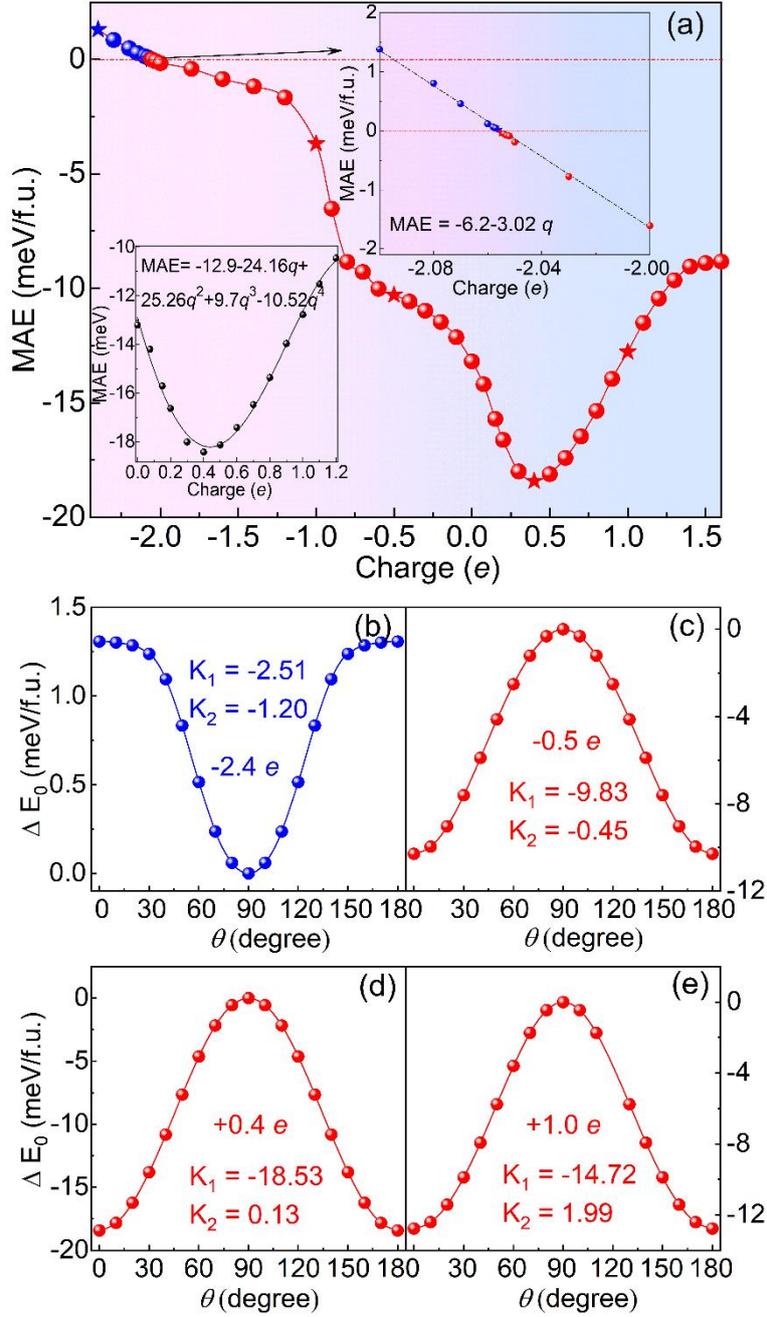

**Figure 8.** (a) The MAE changes with charges doping. The upper right and lower left insets show the MAE changes with the charges of [2.10, 2.00] [0, 1.2] $e$, respectively. (b-e) The energy indicated by the dashed lines changes with θ. The energy varies as a function of polar angle θ of magnetization for $Mn_2Ge_2Te_6$. The energy varies with different doped charges of (b) -2.4, (c) -0.5, (d) +0.4, and (e) +1.0 $e$, respectively.

In order to clarify the change of MAE with charges doping, the tight-binding and second-order perturbation theory are adopted.[61] According to the canonical formulation,[61] MAE of each atom could be calculated, using following equation:

$$MAE_i = \left[ \int E_f (E - E_F)[n_i^{[100]}(E) - n_i^{[001]}(E)] \right] \quad (13)$$

where $MAE_i$ represents the MAE of *i*th atom. $n_i^{[100]}(E)$ and $n_i^{[001]}(E)]$ are the DOS of the *i*th atom with EA along [100] and [001] directions, respectively. The Mn$_2$Ge$_2$Te$_6$ has $D_{3d}$ group, and the energies with EA along [100] and [001] directions are the same.[35] And more detail could be found in Figure S2b. Therefore, only [100] direction is considered here. And total MAE could be rewritten as the sum of $MAE_i$: $MAE_{tot} = \sum_i MAE_i$. According to the second-order perturbation theory,[62] MAE could be gotten by the sum of the following terms:

$$\Delta E^{--} = E_x^{--} - E_z^{--} = \xi^2 \sum_{o^+, u^-} (|<o^-|L_z|u^-|^2 - |<o^-|L_x|u^->|^2)/(E_u^- - E_o^-) \quad (14)$$

$$\Delta E^{-+} = E_x^{+-} - E_z^{+-} = \xi^2 \sum_{o^+, u^-} (|<o^+|L_z|u^-|^2 - |<o^+|L_x|u^->|^2)/(E_u^- - E_o^-) \quad (15)$$

where + and − represent spin-α and spin-β states, and $\xi$, $L_x$, $L_z$ are the SOC constant, angular momentum operators along [100] and [001] directions, respectively. *u*, and *o* represent occupied and unoccupied states, and $E_o$, $E_u$ represent energy of occupied and unoccupied states, respectively. It could be concluded that MAE is mainly dominated by the spin-orbital matrix elements and energy difference. According to the eq 13, the MAE is related with the intensity of DOS. The matrix element differences $|<o^-|L_z|u^->|^2 - |<o^-|L_x|u^->|^2$ and $|<o^+|L_z|u^->|^2 - |<o^+|L_x|u^->|^2$ for *d* and *p* orbitals are calculated, shown in Table 1 and Table 2, respectively. To further interpret MAE changes with charges doping, the atom-orbital-resolved MAE is also calculated, shown in Figure 9 a-i. And it can be concluded that MAE partially come from Mn (Figure 9 a-c) and Ge atoms' contribution (Figure 9 d-f), while it mainly comes from Te atoms (Figure 9 g-i). The orbital-resolved MAE of neutral $Mn_2Ge_2Te_6$ is shown in Figure 9 a, d, g. The total MAE is -13.20 meV/f.u., and Te atoms contribute -11.95 meV. The hybridization between $d_{yz}$ and $d_{z^2}$ of Mn's orbital make positive contribution to MAE, which corresponds to the matrix differences 3 for *d* orbitals, shown in Table 1. The hybridization between $d_{xy}$ and $d_{x^2-y^2}$ orbitals makes negative contribution to MAE, which corresponds to the matrix differences -4 for *d* orbitals. Ge's contribution could be negligible, compared with Te atoms. The hybridization between Te's spin-β occupied $p_y$ and spin-β occupied $p_z$ orbitals is benefit to the in-plane magnetic anisotropy (IMA) (negative value), which corresponds to the matrix differences -1 for *p* orbitals. While the hybridization between occupied spin-β $p_z$ orbitals and unoccupied spin-β $p_x$ orbitals make contribution to PMA (positive value), which corresponds to the matrix 1 for *p* orbitals, shown in Table 2.

**Table 1.** The matrix differences for *d* orbitals between magnetization along [001] and [100] directions in eq 14 and eq 15.

| | $o^+$ | | | | | $o^-$ | | | | |
|---|---|---|---|---|---|---|---|---|---|---|
| $u^-$ | $d_{xy}$ | $d_{yz}$ | $d_{z^2}$ | $d_{xz}$ | $d_{x^2-y^2}$ | $d_{xy}$ | $d_{yz}$ | $d_{z^2}$ | $d_{xz}$ | $d_{x^2-y^2}$ |
| $d_{xy}$ | 0 | 0 | 0 | 1 | -4 | 0 | 0 | 0 | -1 | 4 |
| $d_{yz}$ | 0 | 0 | 3 | -1 | 1 | 0 | 0 | -3 | 1 | -1 |
| $d_{z^2}$ | 0 | 3 | 0 | 0 | 0 | 0 | -3 | 0 | 0 | 0 |
| $d_{xz}$ | 1 | -1 | 0 | 0 | 0 | -1 | 1 | 0 | 0 | 0 |
| $d_{x^2-y^2}$ | -4 | 1 | 0 | 0 | 0 | 4 | -1 | 0 | 0 | 0 |

When -0.5 *e* electrons dope $Mn_2Ge_2Te_6$, the orbital-resolved MAE of $Mn_2Ge_2Te_6$ is shown in Figure S12a-h. The hybridization between Mn's *d*-orbital is similar with neutral $Mn_2Ge_2Te_6$, shown in Figure S12b. However, the hybridization between Te's *p* orbitals changes. Specially, the hybridization between Te's $p_x$ and $p_z$ orbitals, $p_y$ and $p_z$ orbitals are weakened, shown in Figure S12h, whose contribution to total MAE are also decreased to 0.81 and -12.9 meV, respectively. While the contribution of hybridization between $p_y$ and $p_x$ orbitals is enhanced to 2.00 meV. Eventually, the MAE is decreased to -10.29 meV/f.u., shown in Figure 8c and Figure S12 b, e, h. For -1.0 *e* electron doping, the contribution of hybridization between Te's $p_y$ and $p_z$ orbitals is further decreased to -2.86 meV, shown in Figure S12g. As -2.0 *e* electrons are doped, the orbital hybridization between Te's $p_y$ and $p_z$ is further weakened, and total MAE is decreased to -0.161 meV/f.u.. When -2.05, -2.053 *e* electrons are further injected, the corresponding MAEs are -0.019, -0.0077 meV/f.u., as the hybridization between Te's $p_y$ and $p_z$ orbitals is further weakened. When -2.06, -2.08, -2.2 *e* electrons are doped, the corresponding MAEs are 0.012, 0.080, 0.49 meV/f.u., shown in Figure S11 g-i. For -2.4 *e* electrons doping, the occupied $p_y$ is with a bigger energy difference than without doping (Figure S13a), shown in Figure S13b. Therefore, the PMA energy between $p_y$ and $p_z$ hybridization decreases, shown in Figure 9h. As a result, the MAE decreases, according to the eq 15.

**Table 2.** The matrix differences for *p* orbitals between EA along [001] and [100] directions in eq 14 and eq 15.

| $u^-$ | $o^+$ | | | $o^-$ | | |
|---|---|---|---|---|---|---|
|  | $p_y$ | $p_z$ | $p_x$ | $p_y$ | $p_z$ | $p_x$ |
| $p_y$ | 0 | 1 | -1 | 0 | -1 | 1 |
| $p_z$ | 1 | 0 | 0 | -1 | 0 | 0 |
| $p_x$ | -1 | 0 | 0 | 1 | 0 | 0 |

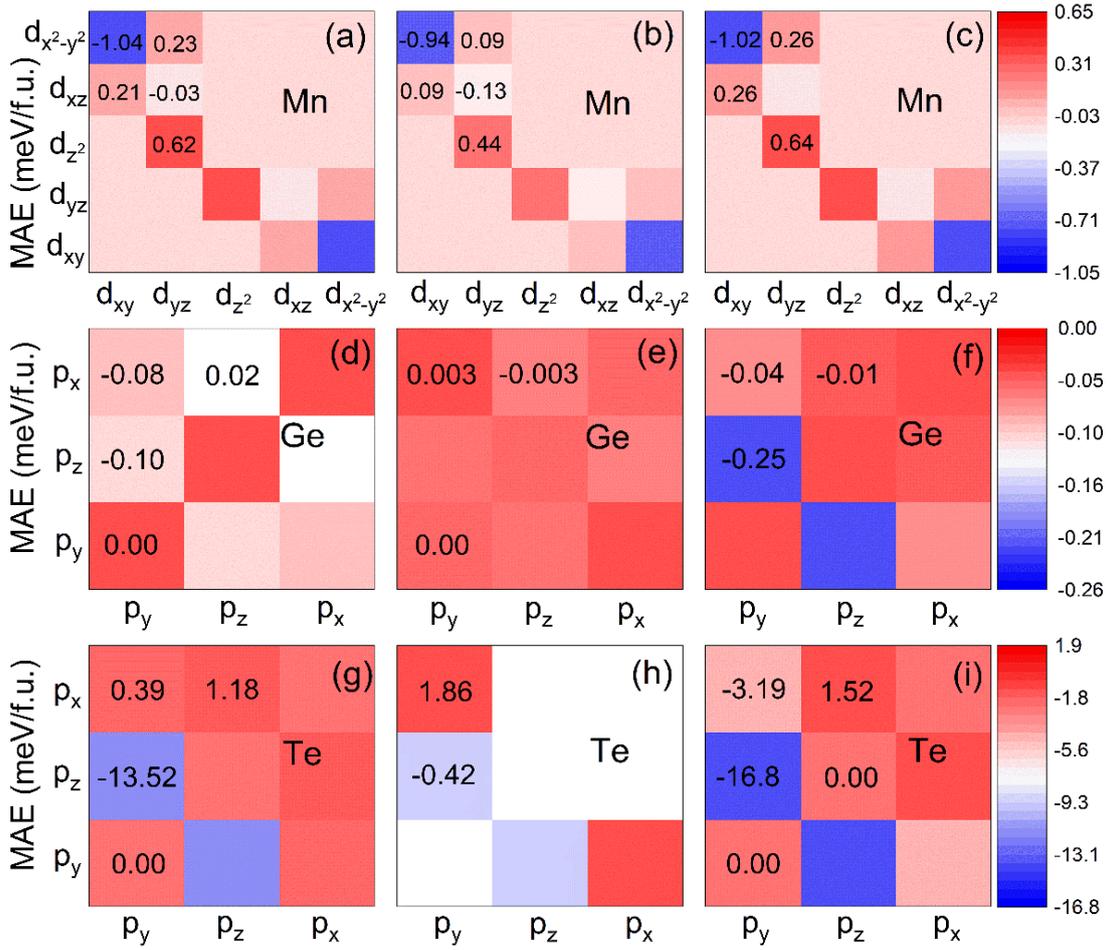

**Figure 9.** Orbital-resolved MAE of $Mn_2Ge_2Te_6$ doped with charges of 0.0, -2.4, and 0.4 $e$, respectively. The orbital-resolved MAE of $Mn_2Ge_2Te_6$ is doped with (a, d, g) 0.0 $e$, (b, e, h) -2.4 $e$, and (c, f, i) +0.4 $e$ charges, respectively.

For the hole doping, the MAEs firstly increase ($q < +0.4\ e$), and then decrease, shown in Figure 8 a, d, e, and Figure S11, 12, 14, respectively. When +0.2 $e$ hole is doped, the hybridization between Te's $p_y$ and $p_z$ orbitals is enhanced, shown in Figure S14a. And the corresponding MAE contributed by Te's atoms is increased to -16.23 meV. When +0.4 $e$ hole is further doped, the MAE contributed by hybridization between Te's $p_y$ and $p_z$ orbitals, $p_y$ and $p_x$ orbitals are increased to -16.76, -3.19 meV, shown in Figure 9i. Compared with undoped $Mn_2Ge_2Te_6$ (Figure S13a), the occupied $p_y$ is closer to the Fermi-level with a smaller energy difference, shown in Figure S13c. Therefore, the PMA energy between $p_y$ and $p_z$ hybridization increases, shown in Figure 9i. For + 1.0 $e$ and +1.4 $e$ holes doping, the MAE contributed by hybridization between $p_y$ and $p_z$ orbitals is decreased to -9.76, -8.67 meV, shown in Figure S12i, S14c, d, respectively. And the total MAEs are -12.77, -9.04 meV, respectively. In a word, the total MAE are dominated by the hybridization between Te's $p_y$ and $p_z$ orbitals.

## 3. CONCLUSIONS

In summary, we have predicted and investigated magnetic and electronic properties of $Mn_2Ge_2Te_6$ ML with PSO method and DFT. We have found intrinsic ferromagnetism in $Mn_2Ge_2Te_6$ ML. All $Mn_2Ge_2Te_6$ ML shows intrinsic FM order, and the ferromagnetism comes from the superexchange interaction between Mn and Te atoms, whose bond angle is close to 90°. $Mn_2Ge_2Te_6$ ML have higher $T_c$ of 316 K. $Mn_2Ge_2Te_6$ is HM with gap of 1.462 eV for spin-β electrons. The corresponding $J_1$, $J_2$ and $J_3$ of $Mn_2Ge_2Te_6$ ML is 13.6, 7.5, and 12.0 meV, respectively. $Mn_2Ge_2Te_6$ ML shows IMA, and corresponding MAE is -13.20 meV/f.u.. $Mn_2Ge_2Te_6$ shows good dynamical and thermal stability. The carriers could effectively modulate magnetic moment, magnetic exchange parameter, and MAE. $Mn_2Ge_2Te_6$ ML shows robust ferromagnetism and half-metallicity under charge doping. However, the total magnetic moment linearly changes with the doping charges, and the exchange parameter could be effectively modulated. Moreover, the MAE could be controlled by changing the hybridization between Te's $p_y$ and $p_z$ orbitals. MAE could be boosted to -18.42 meV/f.u. by enhancing orbitals hybridization with holes doping. Meanwhile, $Mn_2Ge_2Te_6$ transforms IMA to PMA by electron doping. Our work represents robust ferromagnetic half-metallic $Mn_2Ge_2Te_6$ ML with high $T_c$, high MAE, and tunable magnetic properties by carriers doping, making it a candidate for the new magnetoelectronics.

## Author Information

Corresponding Author

*E-mail: zyguan@sdu.edu.cn; Tel: +86-0531-88363179; Fax: +86-0531-88363179

## Acknowledgements

We thank Prof. Wenhui Duan, and Xingxing Li for discussion of evaluation of Curie temperature. We thank Prof. Jun Hu for discussion of second-order perturbation theory. We thank Prof. Shuqing Zhang and, Dr. Rui Li for computation of MCA. We thank Dr. Weiyi Wang about matrix calculation. This work was supported by the financial support from the Natural Science Foundation of China (Grant No. 11904203), and the Fundamental Research Funds of Shandong University (Grant No. 2019GN065). The computational resources from Shanghai Supercomputer Center. The scientific calculation in this paper have been performed on the HPC Cloud Platform of Shandong University. The authors are grateful to Beijing PARATERA Tech Corp., Ltd. for the computation resource in the National Supercomputer Center of Guangzhou.

**Conflict of Interest:** The authors declare no competing financial interest.